\documentclass[11pt,preprint]{aastex}   

\def\etal{{\it et~al.~}}
\def\nt{{nonthermal~}}
\def\bsax{{\it BeppoSAX~}}

\def\ginga{{\it Ginga~}}
\def\einstein{{\it Einstein~}}

\def\asca{{\it ASCA~}}
\def\rosat{{\it ROSAT~}}

\def\chandra{{\it Chandra~}}
\def\rxte{{\it RXTE~}}

\def\erg{~{\rm erg~ cm}^{-2}\ {\rm s}^{-1}~}

\def\cts{~{\rm counts~s}$^{-1}$~}

\begin{document}

\newcommand{\lessim}{\ \raise -2.truept\hbox{\rlap{\hbox{$\sim$}}\raise5.truept
    \hbox{$<$}\ }}

\title{Nonthermal hard X--ray excess in the cluster Abell 2256 \\
from two epoch observations}

\author{Roberto Fusco-Femiano$^a$, Raffaella Landi$^{b,c}$, Mauro Orlandini$^b$
\footnote{$^a$Istituto di Astrofisica Spaziale e Fisica Cosmica
(IASF/Roma), INAF, via del Fosso del Cavaliere, I--00133 Roma,
Italy - dario@rm.iasf.cnr.it; $^b$IASF/Bologna, INAF, via Gobetti
101, I--40129 Bologna, Italy - landi@bo.iasf.cnr.it -
orlandini@bo.iasf.cnr.it; $^c$Dipartimento di Fisica, Universita'
di Bologna, Viale C. Berti Pichat 6/2, I--40127 Bologna, Italy}}

\affil{}

\begin{abstract}
After confirmation of the presence of a \nt hard X-ray excess with
respect to the thermal emission in the Coma cluster from two
independent observations, obtained using the Phoswich Detection
System onboard \bsax, we present in this \textit{Letter} also for
Abell 2256 the results of two observations performed with a time
interval of about 2.5 yr. In both spectra a non thermal excess is
present at a confidence level of $\sim 3.3\sigma$ and $\sim
3.7\sigma$, respectively. The combined spectrum obtained by adding
up the two spectra allows to measure an excess at the level of
$\sim$ 4.8$\sigma$ in the 20--80 keV energy range. The \nt X-ray
flux is in agreement with the published value of the first
observation (Fusco-Femiano \etal) and with that measured by a
\textit{Rossi X-Ray Timing Explorer} observation (Rephaeli \&
Gruber).
\end{abstract}


\section{Introduction}

The formation of diffuse radio regions (radio halos or relics)
detected so far in a limited number of clusters of galaxies seems
due to large-scale shocks and turbulence associated to
gravitational mergers of subclusters and groups able to provide
the necessary ingredients, namely, magnetic field amplification
and particle reacceleration (Tribble 1993; Brunetti \etal 2001;
Fujita, Takizawa, \& Sarazin 2003). In particular, the
megaparsec-scale of radio halos or relics combined with the
relatively short radiative lifetimes of the electrons ($\sim 10^8$
yrs) suggests an in-situ electron reacceleration induced by very
recent or current merger events whose link with diffuse radio
emission seems to be evidenced by X-ray observations (Markevitch
\& Vikhlinin 2001; Govoni \etal 2004). The existence of these
radio regions could be related to the origin of \nt hard X-ray
(HXR) emission which has been detected in a few clusters thanks to
the sensitivity and wide spectral coverage of \bsax and
\textit{Rossi X-Ray Timing Explorer} (\rxte).

A \nt HXR excess has been measured in the Coma cluster by \bsax
and \rxte (Fusco-Femiano \etal 1999; Rephaeli, Gruber, \& Blanco
1999) and recently confirmed by combining the spectrum of the
first \bsax/Phoswich Detection System (PDS; Frontera \etal 1997a)
observation with the spectrum obtained with a second deeper
observation (Fusco-Femiano \etal 2004). The results of these
detections have been challenged by the data analysis of Rossetti
\& Molendi (2004; thereafter RM04). The origin of this difference
is currently under investigation. However, in Sec. 2 we examine
the systematic effects reported in their paper and the possible
systematic error in the net count rates, discussed in Nevalainen
\etal (2004; thereafter NE04), due to unresolved point sources
present in the field of view (FOV) of the PDS .

The \nt flux derived from the combined spectrum of the Coma
cluster, $(1.5\pm0.5)\times 10^{-11}\erg$ in the 20--80 keV energy
range, is consistent with the value of $(1.2\pm0.3)\times
10^{-11}\erg$ measured by the \rxte in the same energy band and
confirmed by a second deeper observation (Rephaeli \& Gruber
2002). Non-thermal radiation has been detected also in A2256 by
\bsax and \rxte (Fusco-Femiano \etal 2000; Rephaeli \& Gruber
2003) and A2319 by \rxte (Gruber \& Rephaeli 2002). At a lower
confidence level, with respect to Coma and A2256, \nt HXR
radiation has been detected by \bsax in A754 (Fusco-Femiano \etal
2003). An upper \nt flux limit has been reported in A3667
(Fusco-Femiano \etal 2001), A119 (Fusco-Femiano \etal 2003) and
A2163 (Feretti \etal 2001). A \nt component is detected at a
$\sim$2$\sigma$ level in $\sim$50\% of the non-significantly
AGN-contaminated clusters observed by \bsax (NE04).

The most likely interpretation of the nonthermal HXR radiation is
inverse Compton (IC) emission by the same radio synchrotron
electrons responsible for the extended radio emission present in
all of the mentioned clusters, scattering the cosmic microwave
background (CMB) photons. It is well know that the alternative
interpretation based on \nt bremsstrahlung emission from
suprathermal electrons (Kaastra, Bleeker, \& Mewe 1998; Ensslin,
Lieu, \& Biermann 1999; Sarazin \& Kempener 2000) has remarkable
energetic problems (Petrosian 2002). Another possible explanation
for the detected \nt emission may be due to a significant
contamination by obscured AGNs (Matt \etal 1999; Hasinger \etal
2001). However, the \chandra observatory that is sensitive enough
to probe a significant fraction of the obscured AGN has not
detected such sources in several clusters (e.g. Molnar \etal
2002). Besides, the co-added spectrum of the sample of clusters in
NE04 gives indication for an extended distribution of the \nt
emission against a significant contamination from obscured AGN.

In this Letter, we present the combined PDS spectrum of Abell 2256
obtained by summing the spectrum of a second long \bsax
observation of $\sim$ 300 ks with that of the previous shorter
observation of $\sim$ 130 ks. The two observations both confirm
the presence of \nt HXR radiation from the cluster in excess of
the thermal emission measured by the Medium-Energy
Concentrator/Spectrometer (MECS) in the energy range $\sim$2-10
keV. Throughout this Letter we assume a Hubble constant of $H_o$ =
70~km~s$^{-1}$~Mpc$^{-1}$~$h_{70}$ and $q_0 = 1/2$, so that an
angular distance of $1^{\prime}$ corresponds to 65.7 kpc
($z_{A2256} = 0.0581$; Struble \& Rood 1991). Quoted confidence
intervals are at $90\%$ confidence level (c.l.), if not otherwise
specified.

\section {PDS Data Reduction and Results}

Abell 2256 was observed for the first time in February 1998 and
then in February 1999 for a total of $\sim$130 ks (OBS1) and
re-observed in July 2001 for $\sim$300 ks (OBS2). The pointing
coordinates of \bsax are at J(2000): $\alpha:~17^h~ 03^m~ 58.3^s$;
$\delta:~ +78^{\circ}~ 38'~ 31''$. The total effective exposure
times of the PDS in the two observations were 70.1 ks and 190.3
ks, respectively. The PDS spectra of both the observations were
extracted using the XAS v2.1 package (Chiappetti \& Dal Fiume
1997). The background sampling was performed by making use of the
default rocking law of the two PDS collimators that samples
ON/+OFF, ON/--OFF fields for each collimator with a dwell time of
96 sec (Frontera \etal 1997a). When one collimator is pointing ON
source, the other collimator is pointing toward one of the two OFF
positions. We used the standard procedure to obtain PDS spectra
(Dal Fiume et al. 1997).

For both OBS1 and OBS2 the +OFF and -OFF count rate spectra are
consistent with each other, with differences equal to $(2.7\pm
3.3)\times10^{-2}$ counts s$^{-1}$ and $(3.3\pm 1.8)\times10^{-2}$
counts s$^{-1}$ in the two observations, respectively. The
marginal evidence (below 2$\sigma$) for the presence of
contaminating sources in the +OFF field in OBS2 implies to
consider the average of the two background measurements in order
to have a more conservative determination of the confidence level
of the excess with respect to the thermal emission. The background
level for OBS1 is 19.50$\pm$0.03 counts s$^{-1}$ in the 15-100 keV
energy range, while for OBS2 is 16.68$\pm$0.01 counts
s$^{-1}$.\footnote{The $\sim$20\% variation in the PDS background
is due to the \bsax orbital decay: the lower orbit for OBS2
increased the shielding to ambient particles, therefore lowering
the diffuse background.} We have examined the presence of the
instrumental background residual reported in RM04, obtained by
these authors analysing the spectra of 15 ``blank fields''. This
residual is instead not found by NE04. We started from the
complete sample of 868 PDS pointings with galactic latitude
$|b|>15^\circ$, and selected the 15--100 keV net count spectra for
which there is source detection below 1$\sigma$ (that is, ``blank
fields''). We summed together these spectra (total exposure $\sim$
2.4 Ms) imposing a net exposure greater than 10~ksec, and found a
net count rate of $(0.66\pm 3.29)\times 10^{-3}$ \cts, consistent
with the definition of ``blank field'' (details in Landi 2005). We
have then investigated the possible systematic differences between
the two background fields $\pm$OFF. RM04 analyse 69 observations
chosen only on the base of a long exposure and high galactic
latitude. They find that the mean value $\langle (-OFF) -
(+OFF)\rangle$ is significantly different from zero. Analyzing a
larger sample, NE04 report evidence for a systematic instrumental
residual which cancels out in the standard usage of both offsets.
In agreement with NE04, our analysis on the {\em whole} sample of
PDS observations gives a value of $(5.3\pm 6.3)\times 10^{-3}$
\cts, consistent with no contamination at all. Finally, using our
PDS blank field sample, we examined the possibility of not
significantly detected point source fluctuation ($\sigma_{fluc}$)
between the offsets. The resulting value $\sigma^2_{fluc}=(9.5\pm
10.3)\times 10^{-4}$ (\cts)$^2$ is consistent with that in NE04.
Since the value is also consistent with zero, we assume in the
following that this background component is absent in A2256
observations.

\subsection{Spectral analysis}

In Fig. 1 we report the PDS and MECS spectra obtained by summing
the spectra of the two observations. The PDS background subtracted
combined count rate is 0.154$\pm$0.011 counts s$^{-1}$ in the
15-100 keV energy range at the confidence level of $\sim
14\sigma$. The PDS data of the first observation has been
reanalyzed using the same procedure utilized for the second longer
observation. The MECS spectrum is in the range $\sim$2-10 keV
obtained from a circular region of 8$'$ corresponding to about 0.5
Mpc centered on the primary emission peak. The \rosat PSPC radial
profile indicates that $\sim 70\%$ of the total cluster emission
falls within this radius (see Fusco-Femiano \etal 2000 for details
regarding MECS data reduction, the cross-correlation between MECS
and PDS and results for only OBS1). The total exposure time is
386.4 ks. The spectral analysis of the MECS data alone gives a
temperature of kT = 7.57$^{+0.19}_{-0.14}$ keV using the MEKAL
code on the XSPEC package, while the simultaneous fit to the MECS
and PDS data gives 7.65$\pm$0.17 keV (the MECS and PDS
normalizations are treated as free parameters) . This value of the
temperature is consistent with the measurements of \ginga
(7.32-7.70 keV; Hatsukade 1989) with a field of view comparable to
that of the PDS, \einstein MPC (6.7-8.1 keV; David \etal 1993),
\asca GIS (6.78-7.44 keV; Henriksen 1999) and with the more recent
joint analysis of \asca and \rxte (PCA \& HEXTE) data (7.66$\pm$
0.12 keV; Rephaeli \& Gruber 2003). \chandra has reported several
emission regions giving evidence for the presence of a merger
event (Sun \etal 2002). These detections show temperature
variations with a mean value of 6.7$\pm$0.2 keV in the central
$8'.3$ square. Also the flux of $\sim 5.4\times 10^{-11}\erg$ in
the 2-10 keV energy range measured by \bsax is consistent with
previous observations. The iron abundance is
0.29$^{+0.01}_{-0.03}$, in agreement with the \asca results
(Markevitch \& Vikhlinin 1997).

The presence of an excess with respect to the thermal emission in
the spectrum of A2256 is evidenced a) by the fit to the PDS data
alone in the 15--80 keV energy range with a thermal component that
determines a temperature of 14.6$^{+4.3}_{-3.2}$ keV well above
the average gas temperature given by \ginga in the interval
7.32-7.70 keV (David \etal 1993); b) by the fit to the PDS data
alone with a bremsstrahlung component at the fixed temperature of
7.6 keV, derived by the MECS data alone, that gives an
unacceptable $\chi^2$ value of 3.96 (=31.69/8 d.o.f.). The \nt
nature of this excess results by fitting the PDS data with two
thermal components, one of these at the fixed temperature of 7.6
keV. For the second component we obtain an unrealistic temperature
greater than $\sim$30 keV. In the joint MECS\&PDS data analysis
the introduction of a second \nt component, modelled as a power
law, allows to obtain an improvement of the $\chi^2$ value that is
significant at more than the 99.99\% confidence level, according
to the \textit{F}-test (198.51/[175 d.o.f.] vs 176.96/[173
d.o.f.]). The presence of the \nt component has the effect of
slightly decreasing the best-fit value of the temperature
(7.32$\pm$0.18 keV) with respect to the temperature obtained by
considering only the MECS data. The confidence contours of the
parameter \textit{kT} and photon spectral index
(\textit{$\alpha_X$}) show that, at a 90\% confidence level, the
temperature is well determined, 7.1-7.7 keV, while
\textit{$\alpha_X$} describes a large interval of 0.3-1.8 (see
Fig. 2). The contribution of the \nt component to the thermal flux
in the 2-10 keV energy range is $\lesssim 10\%$ for
\textit{$\alpha_X$} $\leq$1.8. In the 20--80 keV energy band the
excess with respect to the thermal component derived from the MECS
and PDS data is at level of $\sim 4.8\sigma$ (see Table 1)
considering the average of the two background measurements. The
uncertainties in the thermal flux prediction, reported in Table 1,
are obtained considering the variations of the gas temperature in
the 90\% c.l. range that imply variations of $\sim 0.1\sigma$ for
the level of the \nt excess. The flux of the \nt component,
$8.9^{+4.0}_{-3.6}\times 10^{-12}\erg$ in the 20-80 keV energy
range, is obtained by fixing the photon spectral index
(\textit{$\alpha_X$}) at the best-fit value of $\sim$1.5. The flux
error is derived by varying the PDS normalization within the 90\%
c.l. range. The two separate observations OBS1 and OBS2 show an
hard excess of $\sim$3.3$\sigma$ and $\sim$3.7$\sigma$,
respectively, and the \nt fluxes are consistent with each other
(see Table 1). The difference in the confidence level of the
excess between that shown here for OBS1 and the one presented in
Fusco-Femiano \etal (2000) is mainly due to a more conservative
procedure for the background subtraction and for the bad events
rejection of the PDS data (details in Landi 2005).

\section{Discussion}

The combined spectrum of two \bsax observations performed with a
time interval of $\sim$2.5 yr confirms the presence in Abell 2256
of a \nt HXR component in excess of the thermal emission at a
$\sim$4.8$\sigma$ confidence level. The \nt flux of $\sim 9\times
10^{-12}\erg$ in the energy range 20-80 keV is consistent with the
published value of the first \bsax observation $\sim 1.2\times
10^{-11}\erg$ (Fusco-Femiano \etal 2000). The joint data analysis
of \rxte/PCA\&HEXTE and \asca/GIS\&SIS observations (Rephaeli \&
Gruber 2003) yields evidence for two components in the spectrum.
The secondary component can be either thermal with a low gas
temperature, kT$\sim$ 1.5 keV, or power law with photon index
$\sim$ 2.2 with a \nt flux in the 90\% confidence error interval
of $(0.3-10)\times 10^{-12}\erg$ consistent with our detection.
The confidence contours of two thermal components performed using
the MECS and PDS data (see Fig. 3) confirm the \nt origin for the
second component in the spectrum of A2256. At the 90\% of
confidence the temperature value of the secondary feature must be
greater than $\sim$20 keV. The hottest regions of the cluster
shown by the \chandra temperature map are at the level of $\sim$ 9
keV (Sun \etal 2002). The combined spectrum confirms also the
upper limit of 1.8 for the HXR photon index reported in the first
data analysis of OBS1 (Fusco-Femiano \etal 2000).

As stated in the Introduction, we cannot exclude a significant
contamination by obscured sources located in the PDS FOV able to
simulate the \nt flux detected in Coma and A2256, but so far the
observations seem not to support this interpretation. The most
likely is IC emission by the relativistic electrons responsible
for the radio diffuse emission present in both the clusters. An
extended radio emission (halo) permeates the center of A2256 with
a steep radio spectral index of $\sim$ 1.8-2.0 (Bridle \& Fomalont
1976; Rengelink \etal 1997). This region could be responsible for
the second component that was noted by Markevitch \& Vikhlinin
(1997) in their spectral analysis of the \asca data in the central
r=3$'$ spherical bin. Their best-fit favors a power-law model with
a photon index of 2.4$\pm$0.3 and therefore a \nt component may be
present in the soft region of the X-ray spectrum (details in
Fusco-Femiano \etal 2000). Instead, the probable source of the
nonthermal HXR emission detected by \bsax, in the framework of the
IC model, may be the large and bright relic in the NW region of
the cluster at a distance of $\sim 7'$ from the cluster center.
The extent of this relic is estimated to be $\sim$1.0$\times$0.6
Mpc with a rather uniform spectral index of 0.8$\pm$0.1. This low
value of the radio spectral index indicates a broad reacceleration
region, probably the result of the ongoing merger event due to a
subcluster shown by the X-ray observations. In particular, the
\chandra X-ray image reported in Fig. 1 of the paper of Sun \etal
(2002) shows that the connection of the NW radio relic with the
ongoing merger event seems to be very likely. The merger is
probably still at the early stage (Briel \& Henry 1994) that seems
to be confirmed by the temperature variations across the cluster
that are not so strong as those expected in a major merger. This
phase of the merger event may explain the \nt HXR and radio
synchrotron emissions that are expected only in clusters with
recent or current injection of relativistic electrons because of
their rather short lifetimes due to radiative losses. In the
hypothesis that the NW relic is the source of the \nt HXR flux,
the \bsax observations allow to derive an uniform magnetic field
of $\sim 0.05\mu G$ ($\alpha_R$ = 0.8) in the radio region.
Instead, in the central radio halo a uniform magnetic field of
$\sim 0.5\mu G$ ($\alpha_R$ = 1.8) is derived, assuming a
reasonable contribution of $\sim 5-10\%$ by the power-law
component detected by \asca to the total X-ray flux in the 2-10
keV energy range (Markevitch \& Vikhlinin 1997). A further
evidence that the origin of the \nt HXR radiation detected by
\bsax is probably not due to the central radio halo seems to be
given by the inconsistence between the observed radio index
($\alpha_R$=1.8), that implies a photon index $\alpha_X =
1+\alpha_R$ = 2.8, and the upper limit of the \nt HXR slope of 1.8
(see Fig. 2).

\section{Acknowledgments}

We wish to thank L.Feretti for critical reading of the paper,
F.Frontera for the contribution to the PDS data analysis and the
referee J.Nevalainen for the very useful suggestions.

\clearpage

\begin{deluxetable}{cccccccc}
\rotate
\tablecolumns{8} 
\tablewidth{0pc} 
\tablecaption{Nonthermal HXR excess in the 20--80 keV energy range\label{tab1}}
\tablehead{%
\colhead{Observation} & \colhead{Epoch} & 
\colhead{PDS exposure} & \colhead{T\tablenotemark{a}} & 
\colhead{Observed rate} & \colhead{Predicted rate} & 
\colhead{Excess} & \colhead{Flux\tablenotemark{b}} \\
\colhead{} & \colhead{} & \colhead{(ks)} & \colhead{(keV)} &
\colhead{(counts s$^{-1}\times 10^{-2}$)} & 
\colhead{(counts s$^{-1}\times 10^{-2}$)} & \colhead{(c.l.)} &
\colhead{($10^{-12}\erg$)} }
\startdata 
OBS1 & Feb 1998/99 & 70.1 & 7.43$^{+0.27}_{-0.21}$ &
$10.630\pm 1.854$\tablenotemark{c} &
4.485$^{+0.195}_{-0.083}$\tablenotemark{c} & 3.3$\sigma$ &
9.5$^{+7.8}_{-5.9}$ \\
OBS2     & Jul 2001 &  190.3 & $7.67\pm 0.21$ & $8.324\pm 1.004$ & 4.649$^{+0.170}_{-0.146}$ & 3.7$\sigma$ & $8.0^{+4.4}_{-4.1}$ \\
Combined &     & 260.4 & $7.65\pm 0.17$ & $8.944\pm 0.888$ & 4.628$^{+0.154}_{-0.126}$ & 4.8$\sigma$ & $8.9^{+4.0}_{-3.6}$ \\
\enddata
\tablecomments{Quoted errors for the observed count rates are at 1
$\sigma$.} \tablenotetext{a}{ Derived by the joint analysis of the
MECS and PDS data.} \tablenotetext{b}{\ A photon index of 1.5 was
used to derive the flux (see text for details).}
\tablenotetext{c}{\ In agreement with the value derived by
Nevalainen \etal (2004).}
\end{deluxetable}

\clearpage

\begin{figure}
\rotatebox{-90}{ \epsscale{0.7} \plotone{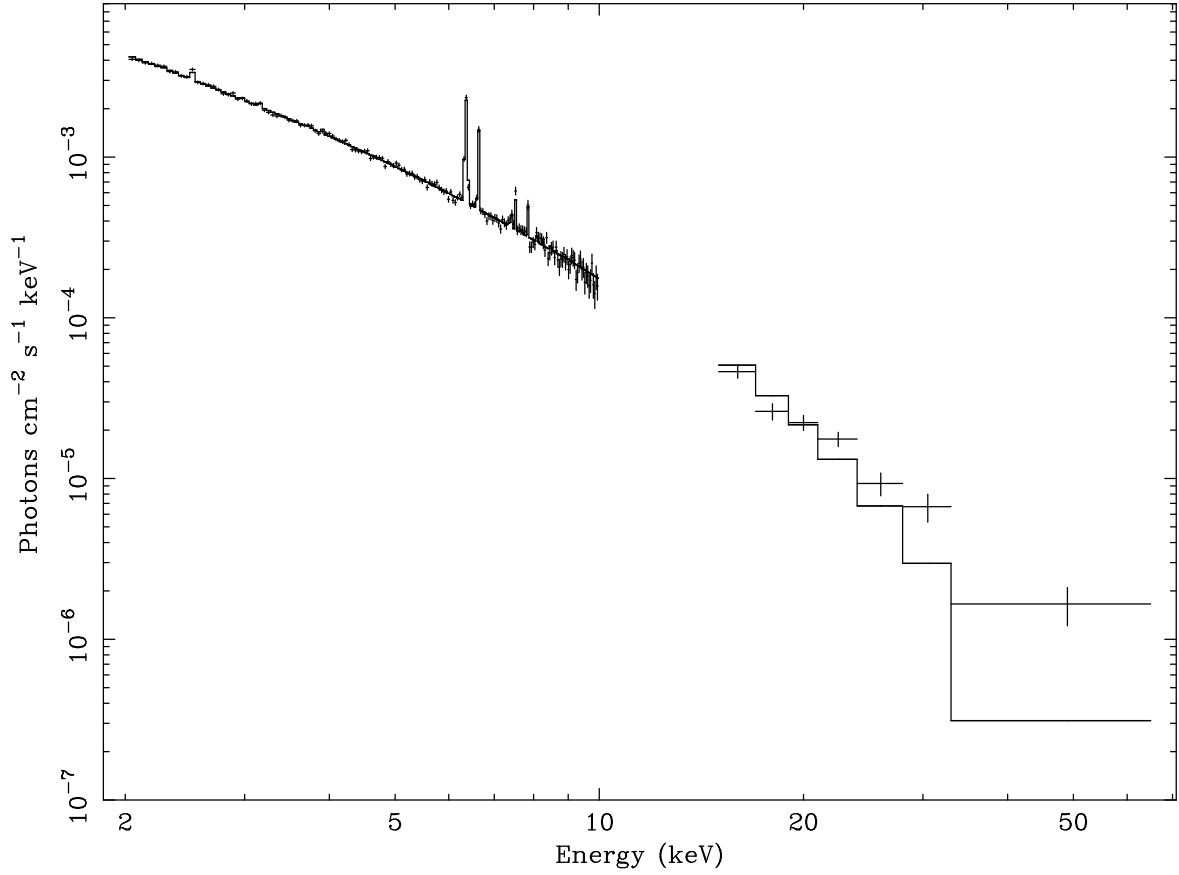}}
\vspace{2cm}
\caption{Abell 2256
--- MECS and PDS data. The continuous line represents a
thermal component (MEKAL code) at the average cluster gas
temperature of 7.65$\pm$0.17 keV in the central circular region of
8$'$ in radius. The errors bars are quoted at the 1$\sigma$
level.\label{fig1}}
\end{figure}

\clearpage

\begin{figure}
\rotatebox{-90}{ \epsscale{0.7} \plotone{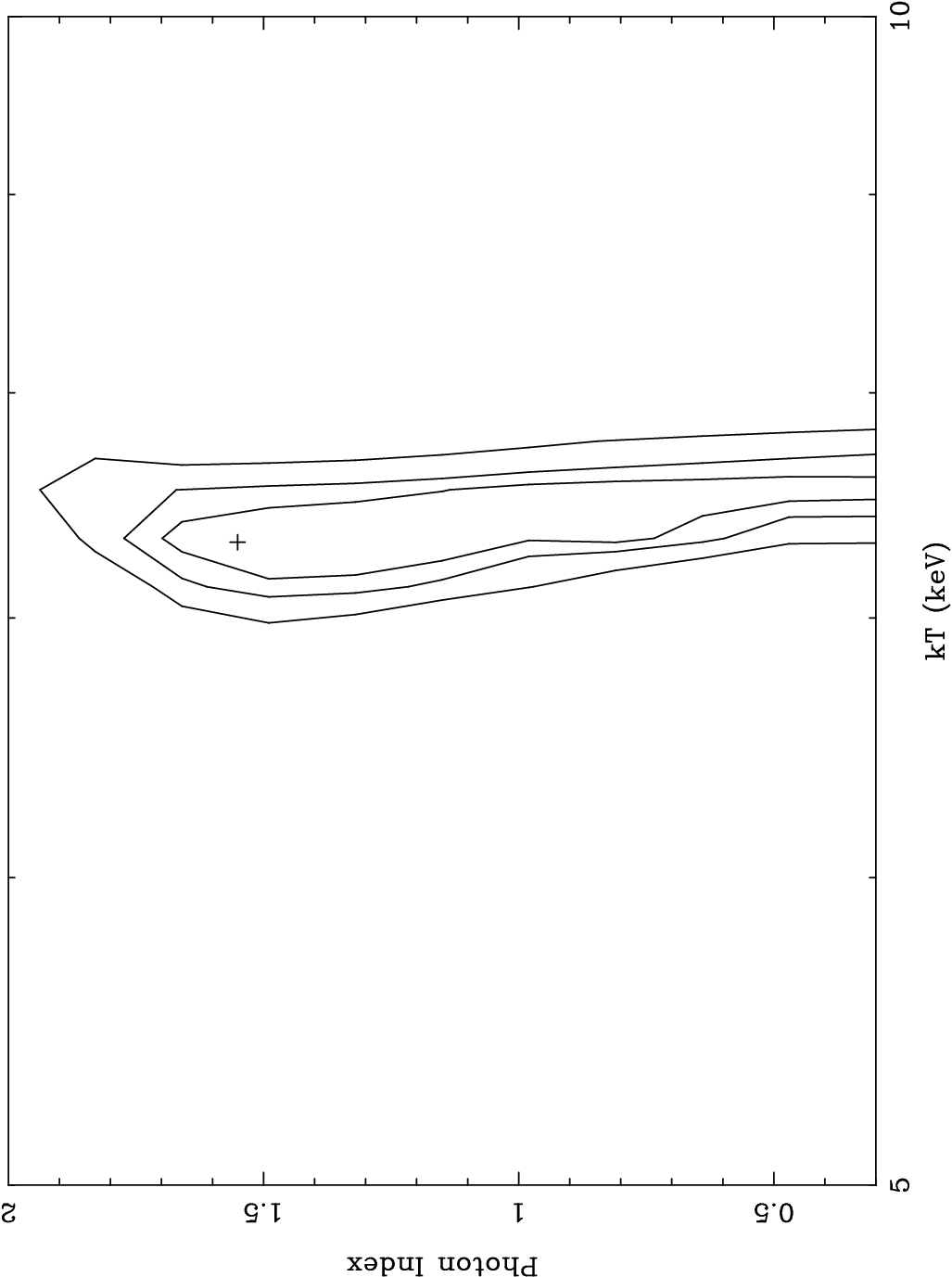}}
\vspace{2cm}
\caption{Confidence contours of a thermal and a \nt component
(MEKAL and POW codes in XSPEC) at the 68\%, 90\% and 99\%
confidence level using the MECS and PDS data.  \label{fig2}}
\end{figure}

\clearpage

\begin{figure}
\rotatebox{-90}{ \epsscale{0.7} \plotone{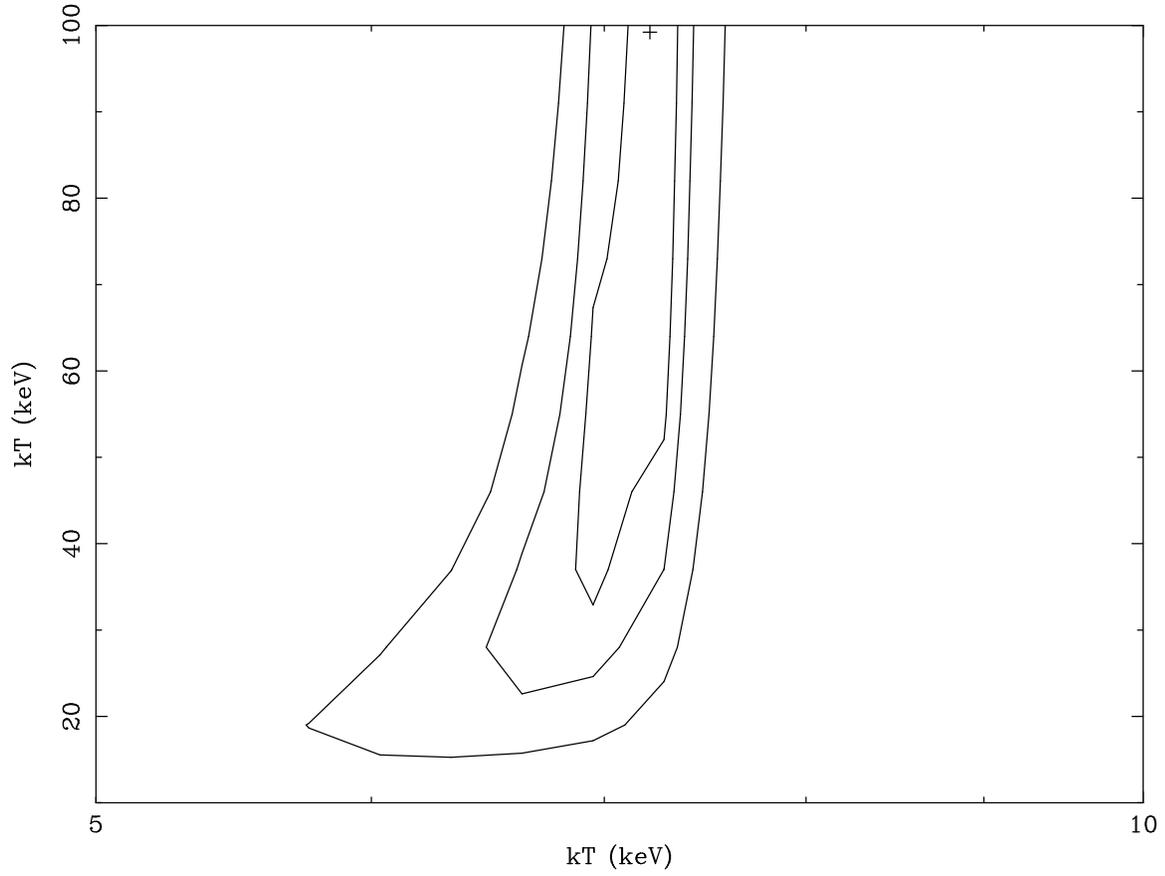}}
\vspace{2cm}
\caption{Confidence contours of two thermal components (MEKAL and
BREM codes in XSPEC) at the 68\%, 90\% and 99\% confidence level
using the MECS and PDS data. \label{fig3}}
\end{figure}

\end{document}